\documentclass[twocolumn,twoside,slac]{revtex4}
\usepackage{graphicx}
\usepackage{fancyhdr}
\begin{document}

\title{Statistical Challenges of Cosmic Microwave Background Analysis
      }

%

\author{Benjamin D.~Wandelt\footnote{NCSA Faculty Fellow}}
\affiliation{University of Illinois at Urbana-Champaign, IL 61801, USA}

\begin{abstract}
The Cosmic Microwave Background (CMB) is an abundant source of cosmological
information. However, this information is encoded in non-trivial ways in a
signal that is difficult to observe. The resulting challenges in extracting this
information from CMB data sets have created a new frontier. In this talk I will
discuss the challenges of CMB data analysis.  I review what cosmological
information is contained in the CMB data and the problem of extracting it.
CMB analyses can be divided into two types: ``canonical'' parameter extraction
which seeks to obtain the best possible estimates of cosmological parameters
within a pre-defined theory space and ``hypothesis testing" which seeks to test
the assumption on which the canonical tests rest. Both of these activities are
fundamentally important. In addition to mining the CMB for cosmological
information cosmologists would like to strengthen the analysis with data from other
cosmologically interesting observations as well as physical constraints. This gives an opportunity   1)
to test the results from these separate probes for concordance and 2) if
concordance is established to sharpen the constraints on theory space by
combining the information from these separate sources.
\end{abstract}

\maketitle
\thispagestyle{fancy}
\section{Overview}

What is cosmic microwave background (CMB) statistics and what  is challenging about it?\footnote{For online material relating to this talk please refer to\\
www-conf.slac.stanford.edu/phystat2003/talks/wandelt/invited/} It involves
estimating the covariance structure of of a spatial random field with
$10^6$--$10^8$ pixels, given only ONE realization of this field. The covariance
matrix of these pixels is a complicated non-linear function of the physical parameters of interest. Of these physical parameters there are between 10 and 20, so even finding the maximum likelihood point is hard---determining and summarizing confidence intervals around the maximum likelihood point is very non-trivial. Cosmologists want to do all this {\em and} have the option of building in exact or approximate physical constraints on relationships between parameters. In addition, since collecting cosmological data is so difficult and expensive we want to combine all available data sets---both to test them for mutual disagreement which might signal new physics, and to improve the parameter inferences. In all of this the quantification of the uncertainties in the results is extremely important---after all the stated significance of our results will either drive or stop theoretical investigations and the design of new observational campaigns.

Before I get on to CMB specifics in section \ref{intro} let me give you the short version of (most)
of this talk for statisticians: ``The CMB is an isotropic (homoschedastic) Gaussian
random field $s$ on the sphere. The desired set of  cosmological parameters
$\Theta=\{\theta_i\, i=1,...,n\}$, are related in a non-linear way to the spatial covariance
structure $S_{ij}\equiv \langle s_i s_j \rangle$ of the field. Observers present us with a sampled, noisy, filtered
and censored/polluted measurement of this field in several 'colors'. The analysis task is two-fold: infer the covariance
structure of the field $s$. Infer the parameters $\Theta$." This is what could be termed ``canonical" CMB analysis.

In this talk I will mainly describe challenges presented by this canonical CMB analysis.  After a brief review of the scientific motivation for studying the CMB in section \ref{intro} I will describe the form of CMB data as well as the current status and prospects of obtaining it in section \ref{data}. Section \ref{inference} then outlines a framework for extracting cosmologically useful information from the data and section \ref{challenges} illuminates some  examples of challenges that arise when implementing this framework. I will  touch on statistical questions concerning ``non-canonical" CMB analysis in section \ref{other} and then conclude in section \ref{conclusions}.

So why are we interested in facing the statistical challenges of CMB analysis?

\section {What can we learn from the CMB?}
\label{intro}
Cosmologists are interested in studying the origins of the physical Universe. In order to do so they have to rely on data. For cosmologists, one of the great practical advantages of Einstein's relativity over Newtonian physics is the fact that we {\em cannot help but look into the past}. Therefore, by observing light that reaches us from farther and farther away, we can study the Universe directly at earlier and earlier times, at least to the extent to which the Universe is transparent to light. Since the early Universe was a hot and opaque plasma we can only see back to the time when the plasma cooled sufficiently (due to the Hubble expansion) to combine into neutral atoms and the mean free time between photons collisions became of order of the present age of the Universe. Photons that we observe today which scattered for the last time in the primordial plasma {\em are} the CMB.

The change from plasma to gas, happened when the Universe was approximately 380,000 years old. CMB photons emitted at this time are therefore the most direct messengers that we can detect today of the conditions present in the Universe shortly after the Big Bang.\footnote{There are other messengers, namely neutrinos and gravitational waves, reaching us from even earlier times but we do not (yet) have the technology to detect them in relevant quantities.} They constitute a pristine snapshot of the infant Universe which provide us with direct cosmological information uncluttered by the complex non-linear physics which led to the formation of stars and galaxies. One of the main attractions of the CMB is the conceptual simplicity with which it can be linked to the global properties of the Universe and the physics which shaped it at or near the Planck scale.

As a simple example, the serendipitous discovery of the CMB by Penzias and Wilson in 1965 showed that the CMB is isotropic to a very high degree. This was one of the key motivations for the development of the inflationary paradigm \cite{Guth,Linde,AlbrechtSteinhardt}. Inflation describes generically the emergence (from the era of quantum gravity) of a large homogeneous and isotropic Universe. Inflation also predicted the spectrum of small metric perturbations from which later structure developed through the gravitational instability (though it was not the only mechanism to do so). The corresponding anisotropies in the CMB were first convincingly detected by the DMR instrument on the COBE satellite in 1992. The rejection of alternative mechanisms for the generation of the primordial spectrum of metric perturbations in favor of inflation was a major advance driven by the measurement of the large angle CMB anisotropy. These developments have led to inflation becoming part of the current cosmological standard model. A very robust prediction of generic model implementations of inflation is the Gaussianity and homogeneity of the resulting perturbations.

Quantitatively, the properties of our Universe are encoded in a set of $n\sim 10-20$  {\em cosmological parameters} $\Theta$ where $n$ depends on the level of detail of the modeling or, commonly, on the specification of theoretical priors which fix some of these parameters to ``reasonable" values. These parameters specify the geometry and average energy density of the Universe, as well as the relative amounts of energy density contributed by the  ingredients of the primordial soup (dark matter, ordinary baryons, neutrinos and dark energy and photons). In addition, the anisotropy carries information about the spectrum of primordial (inflationary) perturbations as well as their type (adiabatic or isocurvature). By combining observations of the anisotropy of both the effective temperature and the polarization of the CMB photons we can infer how transparent the Universe really was for the CMB photons on their way from last scattering to hitting our detectors. This in turn can tell us about the history of star formation.

A very exciting prospect is that by studying the details of CMB polarization we can infer the presence or absence of gravitational waves at the time of last scattering. A detection would offer an indirect view of one of the elusive messengers that started their journey at an even earlier epoch, adding a nearly independent constraint on the properties of the Universe at the Planck scale.

All this information is not encoded  in actual features in the CMB map of (temperature or polarization) anisotropies. In fact in a globally isotropic universe the absolute placement of individual hot and cold spots is devoid of useful information. Information can, however, be stored in the invariants of the photon brightness fluctuations under the group of rotations SO(3). These are the properties of the field that only depend on the {\em relative} angular distance between two points of the field. For a Gaussian field, where 2-point statistics specify all higher order moments, this means that the angular power spectrum coefficients of the anisotropies contain all of the information.

The challenge for theoreticians was then to develop a detailed theory of the angular power spectrum $C_\ell$, as a function of angular wavenumber $\ell$, given the cosmological parameters $\Theta$. While conceptually simple, it required a decade-long intellectual effort to model the relevant physical processes at the required level of precision. As a result, there now exist several Boltzmann codes (e.g. CMBFAST \cite{CMBFAST} or CAMB \cite{CAMB}), which numerically compute  $C_\ell(\Theta)$ to 1\% precision or better. The power spectra $C_\ell(\Theta)$ are sensitive functions of certain combinations of the parameters and weak functions of others (degeneracies). These weakly constrained parameter combinations are referred to as {\em degeneracies}. Within the context of the standard cosmological model, the theory of this dependence is well-understood.

It is clear from the preceding discussion that the CMB is an extremely valuable source of what amounts to ``cosmological gold": information about the physics and the global properties of the early Universe. So what are the observational prospects?

\begin{figure*}
\centering
\includegraphics[width=140mm]{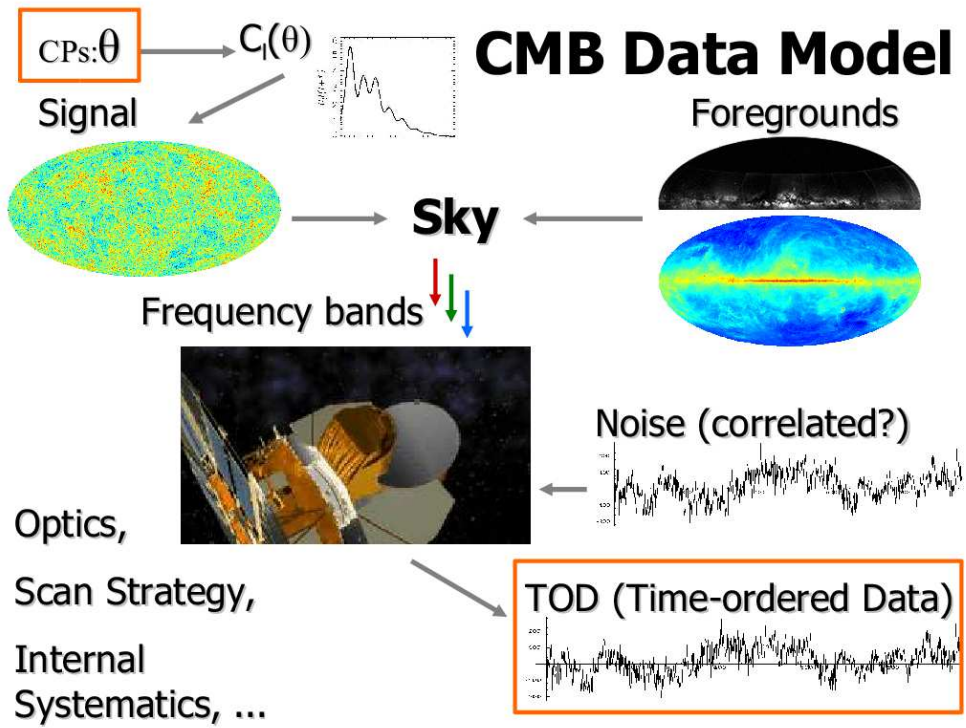}
\caption{A schematic of how the cosmological parameters $\Theta$ (top left) are linked to the time ordered data CMB experiments  actually observe (bottom right). Please see the discussion in the text. } \label{datamodel}
\end{figure*}

\section{Data from CMB Observations}
\label{data}
A major international effort is underway to make high quality observations of the microwave sky using ground-based, balloon borne and space missions \cite{CMBobs}. Space missions have the advantage of being able to scan the whole microwave sky. NASA's ``Wilkinson Microwave Anisotropy Probe" (WMAP) was launched in 2001 with great success and is currently in operation. It reported its first year of data earlier this year \cite{WMAP}.  WMAP will continue to collect data for at least another three  years. In the medium term (ie.~in late 2007) we anticipate the launch of ``Planck," a joint ESA/NASA space mission.\footnote{http://www.rssd.esa.int/index.php?project=PLANCK}  Additional space missions may follow that will focus on measuring the polarization anisotropies in the CMB.
In the meantime ground and balloon-based missions are jostling to accelerate our learning curve by providing maps at high angular resolution on small patches of the sky (up to a few degrees large). As a result,  by the end of this decade we will have a mountain of CMB data.

From this mountain (and of course from the part of it which we have already available today) we would like to extract the cosmological gold. In order to do so we need to understand how the data and the information are related. In Figure~\ref{datamodel} I show a simplified schematic of this relationship and I will now go through the various steps.

Starting with a set of cosmological parameters $\Theta$ we can use a Boltzmann code to compute the power spectrum $C_\ell$. Given this power spectrum and the assumption of Gaussianity and isotropy we have all the information we need to create a statistical realization of a CMB anisotropy map. This is most simply done working in the Fourier representation: we draw the spherical harmonic coefficients $a_{\ell m}$ from uncorrelated Gaussians with zero mean and variance given by $C_\ell$. Then we compute the spherical harmonic transform to obtain
\begin{equation}
T(\hat{n})=\sum_{\ell=2}^{\ell=\infty}\sum_{m=-\ell}^{m=\ell} T_{\ell m} Y_{\ell m}(\hat{n}),
\label{SHT}
\end{equation}
where $ Y_{\ell m}(\hat{n})$ are the spherical harmonics, an orthonormal and complete basis for functions on the sphere. The unit vector $\hat{n}$ points in a direction on the sky.

Unfortunately, the $T_{\ell m}$ or $T(\hat{n})$ cannot be observed directly. We are tied to our location in the Galaxy. There are various sources of copious amounts of microwave radiation in the Galaxy, such as dust and  electrons from winds which are accelerated in the Galactic magnetic field. These foregrounds add to the CMB signal to make our sky.

This sky is then observed in various frequency bands (indicated by the red, green and blue arrows) by a CMB instrument (the figure shows an artist's conception of the WMAP satellite \cite{wmapsite}). This instrument itself has a complicated transfer function: since weight and size constraints force satellite optics to be built compactly the optics are not free from distortion. Microwaves have macroscopic wavelengths and therefore diffract around the edges of the instrument. This leads to sidelobes in the beam maps. The instrument scans the sky in a certain pattern (the ``scan strategy") and internal instrument systematics  are added to the scanned signal. The microwave detectors (either radiometers or bolometers) add noise to generate the time ordered data (TOD).

There are different levels of detail of the CMB data. To give an idea for the orders of magnitude involved, let us go through the sizes of the various objects in Figure~\ref{datamodel} for the Planck experiment.
\begin{itemize}
\item The complete TOD for Planck will take up of order 1 Terabyte (=$10^{12}$ bytes) of storage (without counting house-keeping and pointing data).
\item Each of the 100 detectors (channels) results in a map which is of order 10-100 Megabytes.
\item The channels are grouped into of order 10 frequency bands.
\item The combined maps at these different frequency bands will be combined into maps of the physical components (such as dust, synchrotron, CMB).
\item The CMB power spectrum has a few thousand coefficients $C_\ell$.
\item These power spectrum coefficients are a function of $10-20$ cosmological parameters.
\end{itemize}
Note that there is a trade-off between the level of compression and what assumptions are implemented in the data analysis. Note also, that except for the raw data each of these data products  by themselves mean little unless some means of assessing their statistical uncertainty is provided.

We immediately run into practical problems. For example, if we would like to specify a noise covariance matrix for each combined map at each frequency we would have to specify 10 times $\sim(10^6)^2/2$ elements of a matrix. The necessary storage space of order 10,000 Gigabyte basically precludes practical public distribution of the data. Other ways of specifying the uncertainty must be found.

Before we discuss in detail how cosmological information is encoded in the data, let us comment about an aspect of CMB statistics that is challenging even before there is any data in hand. This aspect is {\em experimental design}. While our experimental colleagues are putting a great deal of valuable thought into their instruments and observational strategies, it is currently still done in an informal way, by ingenuity rather than by formal method. This is partially so because it is hard to define optimality criteria that everyone would agree with and partially because of the immense effort involved to actually carry out the optimization. Evaluating any one proposed design requires many simulations of the full process from observation to analysis---the very process that presents the challenges I am discussing in this talk. Doing this repeatedly to search the space of design parameters for an optimal solution would be an immense task. Asymptotic techniques have been implemented on the basis of the Fisher matrix minimum variance predictions, but it is important to keep in mind that these are lower bounds on the expected variance, assuming a unimodal, Gaussian likelihood shape.

\section{Inference From the Data}
\label{inference}
A stochastic model of the data and the information contained in it can be summarized in the following equations. For simplicity we will limit the discussion to a single channel. The TOD, $d$ say, is modeled as the result of the action of a linear operator $A$ which encodes the optics and scanning strategy of the instrument, on  the sky, made up of signal $s$ and foregrounds $f$:
\begin{equation}
d=A(s+f)+n.
\end{equation}
Our assumptions is that $s$ is an isotropic Gaussian random field with zero mean and power spectrum $C_\ell$. We will take the noise correlations  $N=\langle nn^T\rangle$ and $A$ as given---though one of the statistical challenges of CMB analysis is to relax this assumption to some degree. We will not discuss this further here.

The task is then to extract as much information as possible about the cosmological parameters $\Theta$ from the TOD. To set up this inverse problem we write down Bayes' theorem
\begin{equation}
P(s,C_\ell,f,\Theta|d) P(d) = P(d|f,s,C_\ell,\Theta) P(f,s,C_\ell,\Theta).
\label{bayes}
\end{equation}
In the Bayesian context, solving the inverse problem means exploring and summarizing the posterior density $P(s,C_\ell,f,\Theta|d)$. On the right hand side we can use the conditional independence of the $C_\ell$ and the data given $s$, and the plausible independence of $f$ and $s$ to simplify
\begin{equation}
P(f,s,C_\ell,\Theta)=P(f)P(s|C_\ell)P(C_\ell|\Theta)P(\Theta).
\end{equation}

Traditionally, inference is performed in a linear sequence of steps which are concatenated into a pipeline. At each individual step a likelihood is written for the data in its current representation (e.g. TOD, maps, $C_\ell$) in terms of the parameters describing the next stage of compression. Due to the  complexity of evaluating the likelihoods,  often the likelihood approach is abandoned and approximate, suboptimal but unbiased estimators are constructed.

Within the above framework we can understand each of the steps as a limit of Eq.~\ref{bayes}. For example the ``map-making" step is signal estimation with
\begin{equation}
P(f,s,C_\ell,\Theta) = P(f)P(s|C_\ell)P(C_\ell|\Theta)P(\Theta)= const.
\label{mapprior}
\end{equation}

Similarly, ``Power spectrum estimation" summarizes the marginal posterior $P(C_\ell|d)$, where the $d$ is taken in compressed form as the estimate of the CMB component. The summary proceeds through a maximum likelihood estimate (MLE) and the evaluation of the curvature of $P(C_\ell|d)$ at the MLE. In fact in most practical cases a non-optimal estimator is used due to the computational complexity of evaluating $P(C_\ell|d)$ or its derivatives and hence of computing the MLE.

Finally, ``parameter estimation" signifies the step from the MLE of $C_\ell$ to a density $P(\Theta|d)$. The posterior $P(\Theta|d)$ is usually summarized in terms of the marginalized means and variances of the parameters or in terms of two-dimensional projections or marginalizations through the n-dimensional parameter space.

Recently, we developed MAGIC, a method that shows promise for global inference from the joint posterior \cite{MAGIC}. The method, based on iterative sampling from the joint posterior, exploits the full dependence structure of the different science products from a CMB mission. For example it is useful for the separation of signal and foregrounds if the covariance of the signal is known. The information obtained on $C_\ell$ can be fed back into the CMB map estimate, which would in turn lead to a better estimate of $C_\ell$, etc. But since I discussed this technique in my contributed talk \cite{wandeltcontrib} I will not go into details here.

I will now discuss a selection of the challenges we face in the ``map-making," ``power spectrum estimation" and ``parameter estimation" steps on the way from the TOD to $\Theta$.

\section{Challenges in CMB Analysis}
\label{challenges}
\subsection{Map-making Challenges}

{\bf Pixelizations of the sphere.} A very basic challenge in CMB analysis is the fact that the CMB is a random field on the {\em sphere}. Convenient numerical techniques for storing and manipulating functions on the sphere needed to be developed. For CMB analysis in particular, fast methods for spherical harmonic transforms had to be developed and implemented. Point sets were needed that had good generalized quadrature properties such that discrete numerical approximations to Eq.~\ref{SHT} and its inverse
\begin{equation}
a_{\ell m}=\int d^2\hat n Y_{\ell m}(\hat n) T(\hat n)
\label{ISHT}
\end{equation}
generate accurate results. For local operations such as nearest neighbor searches and multi-resolution work, pixelizations of the sphere that allow hierarchical refinement are useful.

Various pixelizations of the sphere have been proposed \cite{HEALPix,Igloo,GLESP}. Of these, HEALPix features pixels with exactly equal areas throughout, approximate equidistribution of pixel centers over the sphere, simple analytical equations for the pixel boundaries, fast spherical harmonic transforms and very favorable quadrature properties. Due to these features, HEALPix has developed into the standard pixelization for astrophysical all sky maps.

{\bf Beam Deconvolution Map-making.}
It is easy to show that the MLE $\hat m$ for the map $m=s+f$ is the result of maximizing Eq.~\ref{bayes} using Eq.~\ref{mapprior}. This results in
\begin{equation}
m\equiv(A^TN^{-1}A)^{-1}A^TN^{-1}d
\end{equation}
with the associated noise covariance matrix $C_N=<m
m^T>=(A^TN^{-1}A)^{-1}$.

These equations are generally valid whatever the forms of $A$ and $N$. However, the non-trivial structure in the observation matrix $A$ which is a consequence of the unavoidable imperfections in satellite optics makes solving for the map challenging. It is important to deconvolve the beam functions, since the beam distortions and side lobes lead to spurious shadow images of bright foregrounds. These images are spread throughout the map in a complicated way that depends on the scanning strategy. In addition, not accounting for beam imperfections results in distorting the signal itself. Both of these effects can bias not just the map but also the result of the covariance estimation.

For polarization map-making this deconvolution becomes even more important, since the polarization signal is weak and beam asymmetries can introduce spurious polarization signals into the data. Further, if bright foregrounds are significantly polarized, they may induce a significant polarization through their shadow images if beam convolution effects are neglected in map-making. Making high-quality maps of the polarization of the CMB anisotropy is the next frontier in CMB map-making.

To simulate the effects of realistic beams we have to have a general  convolution technique for a beam map with a sky map along the scan path. If implemented as discrete sums over two pixelized maps, rotated to all possible relative orientations, such a technique requires of order $n_p^{2.5}$  operations, where $n_p$ is the number of pixels in the maps. By doing the convolution in spherical harmonic space it was shown in \cite{wandeltgorski} that this can be reduced to $n_p^2$ in the general case and to $n_p^{3/2}$ in interesting limit cases. This fast convolution method was generalized to polarization in \cite{challinor}.

\begin{figure*}
\centering
\includegraphics[width=140mm]{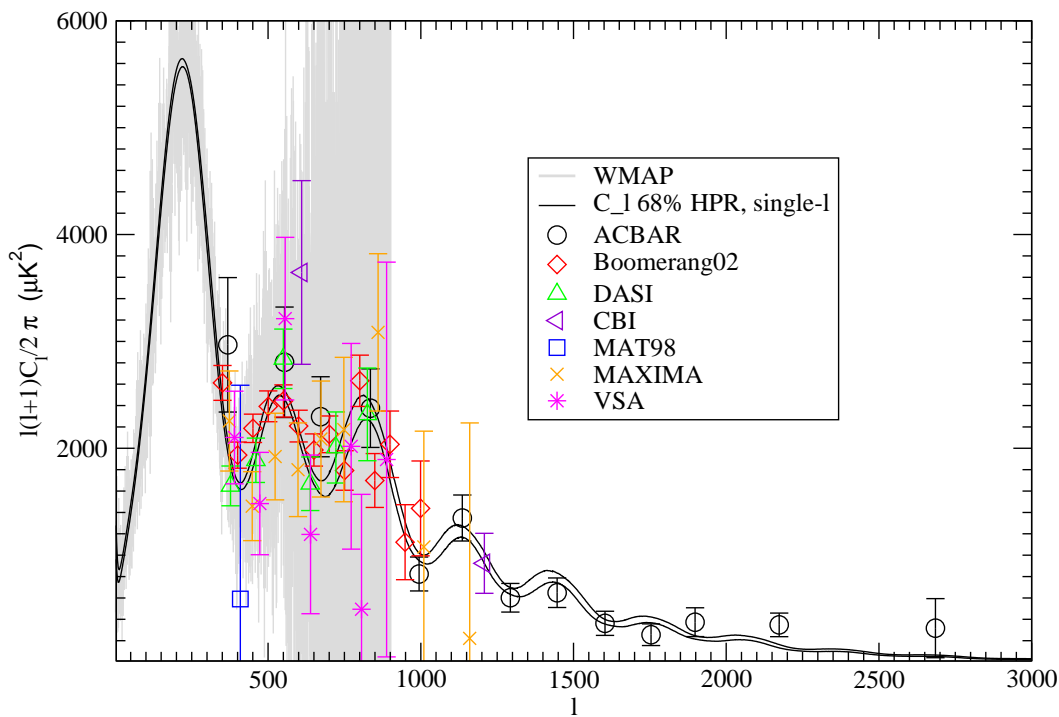}
\caption{Our compilation of all recent CMB power spectrum ($C_\ell$) data (points), including the WMAP data (gray band).  Pre-WMAP data which provide redundant information about the power spectrum at low $\ell$  are omitted. Also shown (as two solid lines) is the 68\% constraint on the power spectrum after implementing a prior which restricts the range of theories to a 10-parameter space of adiabatic inflationary theories (from \cite{HCW}).} \label{compile}
\end{figure*}

\begin{figure}
\centering
\includegraphics[width=75mm]{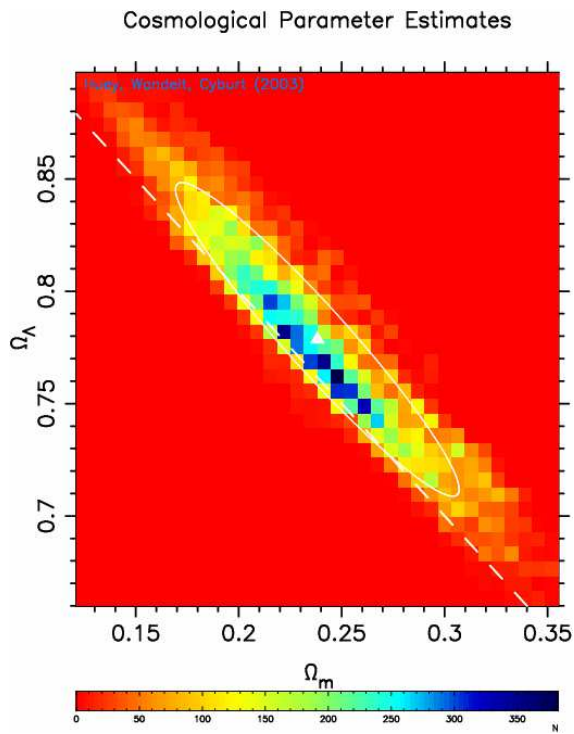}
\caption{An example result of an online exploration of the marginal posterior of the dark matter density ($\Omega_m$) and dark energy density ($\Omega_\Lambda$) in our Universe. All recent CMB data, including the WMAP data, as well as the Hubble Space Telescope key project results and the Supernova cosmology project were included to obtain these constraints. Points in the figure are colored (from red to blue) according to how well the parameter combination at that point agrees with the data.} \label{OLOm}
\end{figure}

{\bf Component separation}. Once a map of the sky is made from each channel, these maps can be compressed losslessly into maps at each frequency. A great deal of work in the field has gone into devising methods for then obtaining an estimate of the CMB sky from these foreground contaminated maps at each frequency band, both for temperature and for polarization. We can either choose to model the foregrounds physically (e.g. \cite{Stolyarov,WMAPFG}) or we can attempt ``blind separation," by defining an algorithm for automatic detection of different components in the maps, e.g. based on the statistical independence of these components (e.g. \cite{Maino}).

{\bf Lensing}. The CMB has traveled past cosmological mass concentrations which perturb the photon geodesics. This distortion contains valuable and complementary information about cosmological parameters. At the same time it mixes the polarization modes in the CMB data, contaminating the primary signal for the detection of the primordial gravitational wave background. Methods need to be devised that can measure this distortion and extract the information contained in it. It is intriguing that \cite{HirataSeljak} find that exact techniques have significant advantages over approximate, quadratic estimators for the reconstruction of B polarization maps from lensed CMB.

\subsection{Power Spectrum Estimation}
{\bf The Computational Problem.} For perfect (all-sky, pure signal, no noise) data, power spectrum estimation would be easy. Just compute the spherical harmonic transform, Eq.~\ref{ISHT}. Then the estimator
\begin{equation}
\hat{C_\ell}=\frac{\sum_m |a_{\ell }|^2}{2\ell+1}
\end{equation}
is the MLE for ${C_\ell}$. It is also easy to evaluate and explore the perfect data posterior to quantify the uncertainty in the estimates.
However, in the general case we would like to evaluate $P(S(C_\ell)|d,N)$ which is the result of integrating out (``marginalizing over'') $s$ in the joint posterior. We obtain\footnote{We use $G(x,X)$ as a shorthand for the
multivariate Gaussian density
\begin{equation}G(x,X)=\frac{1}{\sqrt{|{2\pi X}|}}\exp\left({-\frac12 x^T X^{-1} x}\right).
\end{equation}
}
\begin{equation}
P(S(C_\ell)|d,N)=G(m,S(C_\ell)+C_N).
\label{PofSgivend}
\end{equation}
Here $S$ is the signal covariance matrix, parameterized by the $C_\ell$. Since $S+C_N$ is not a sparse matrix and since the determinant in the Gaussian depends on $S$, evaluating $P(S(C_\ell)|d,N)$ as function of the $C_\ell$ costs of order $n_p^3$ operations.

For Planck, $n_p\sim 10^7$ so if the constant factor in the scaling law was $1$ (an unrealistic underestimate) this would mean $10^{21}$ operations for one likelihood evaluation. For a 10 GFLOP CPU this means 1000s of CPU years of computation.

Various approaches have been suggested to to counter this challenge. They can be broadly divided into two classes: 1) specialized exact (maximum likelihood estimation) algorithms exploit advantageous symmetries in the observational strategy of a CMB mission \cite{OSH,WH} to reduce the computational scaling from order $n_p^3$ to order $n_p^2$, and 2) approximate algorithms which filter unwanted properties of the data and simply compute the power spectra on the filtered and incomplete data, and then de-bias the results using Monte Carlo simulations after the fact\cite{master}. This second class of algorithms is known as pseudo-$C_\ell$ algorithms\cite{WHG} and has gained a great deal of popularity in recent analyses of CMB data, including the WMAP data \cite{WMAPCl}.

{\bf Cosmic Variance.} Aside from computational problems there is an interesting conceptual problem with CMB power spectrum analysis: the fact that we only have one sky.

This fact induces a fundamental limitation to how well we will be able to constrain cosmological parameter estimates from the CMB. The usual way of phrasing this limitation is in terms of cosmic variance.

Essentially, power spectrum estimation is variance estimation. For perfect data the solution that maximizes Eq.~\ref{PofSgivend} is
\begin{equation}
\hat{C_\ell}=\frac{\sum_m |a_{\ell }|^2}{2\ell+1}.
\label{cosvar}
\end{equation}
So we are estimating the variance of the spherical harmonic coefficients.
But, for example, for $\ell=2$ there are only 5 such coefficients and a variance estimate from 5 numbers is statistically uncertain. So ``cosmic variance" is expression of the fundamental limit to the precision of any measurement of the $C_\ell$  caused by the fact that the sphere is a bounded space and the fact that causality will not allow us to observe independent patches of the universe.

This fundamental limit to our knowledge provides a powerful motivation to do the best possible job in analyzing cosmological data.

\subsection{Parameter Estimation}

{\bf Techniques.} The problem of parameter estimation is challenging since we need to explore the posterior density $P(\Theta|d)$ which varies over $\sim 10-20$ dimensions. Various techniques have been used to do this. For smaller number of dimensions (up to about 5) gridding techniques have worked well. However, the current state of the art is to use Markov Chain Monte Carlo methods \cite{Knox} such as the Metropolis Hastings algorithm to sample from $P(\Theta|d)$ and to then base inferences on summaries of the posterior density computed from the sampled representation.

The question how to implement physical priors and constraints was one theme that was discussed at this conference. Within the Bayesian framework there exists a unique prescription for applying physical constraints through the specification of informative priors.
The first example of a CMB parameter estimation which can be explored interactively online is the Cosmic Concordance Project \cite{HWC}. This compiles data from several recent CMB observations and combines them with the user's choice of other, non-CMB experiments and physical priors. The result is displayed as the 2D marginal posterior density for any 2 parameters chosen by the user. An example is displayed in figure \ref{OLOm}.
I invite you to have a look at our prototype implementation at http://galadriel.astro.uiuc.edu/ccp.

 A first scientific result from this project was a measurement of the fraction of $^4$He in the Universe, both from CMB data alone and in the combination of standard Big Bang Nucleosynthesis with the CMB data. Since standard BBN links the primordial  $^4$He abundance to the baryon to photon ratio which is determined exquisitely well by CMB data, we obtain the most precise measurement of the primordial  $^4$He abundance to date \cite{HCW}.

Several open questions remain to be addressed. What number $n$ of parameters is needed to fit the data? Which selection of parameters from the full set should we use in the analysis? These questions are currently handled on the basis of personal preference of the authors or computational convenience---a statistically motivated procedure for letting the data selecting certain parameters has  not yet been implemented.

Operationally, it is very computationally expensive to sample from a posterior with 10 parameters or more, because each likelihood evaluation requires running a Boltzmann code which computes $C(\Theta)_\ell$. In many dimensions the Metropolis sampler produces correlated samples (regardless of whether the target density is correlated or not---the correlations come from the sampler taking small steps through the many dimensional space). New numerical techniques for evaluating and sampling from the likelihood in high-dimensional spaces are needed.

 {\bf Beating cosmic variance.} In Figure~\ref{compile} we show another application of parameter estimation techniques which is relevant to several themes we touched on in this talk and at this conference. The figure shows our compilation of CMB data and the mean and $\pm 68$\% errors on the range of power spectra which are contained in our 10 parameter fits. In other words, we have implemented the optimal non-linear filter for the $C_\ell$ if the Universe is really described by our 10 parameter model. The physical prior has reduced the cosmic variance error bars far below the limit set by Eq.~\ref{cosvar}. Since we are using the physical prior that the CMB power spectrum is the result of plasma oscillations in the primordial photon baryon fluid the resulting smoothness of the power spectrum is used in constructing the estimate.

\section{Testing the Assumptions: Challenges of Non-canonical CMB Analysis}
\label{other}
In addition to the canonical analyses outlined above it is fundamentally important to test the assumptions on which these analyses rest. I will very briefly mention two of them here.

Is the CMB signal really Gaussian? Is it isotropic? These questions touch simultaneously on the issue of hypothesis testing, as well as model selection. What is the evidence in the data for the assumptions of isotropy and Gaussianity on which canonical analyses of the CMB are built? Can a goodness of fit criterion be defined which allows assessing whether the standard model of cosmology is a complete description of the CMB data?

The idea of testing the goodness of fit can be extended to cross tests of CMB data with other cosmological data. Ultimately the goal of the Cosmic Concordance Project is to allow users to select various data sets and explore interactively whether the parameter constraints from various data sets are compatible with each other. If this agreement is established the the constraints can be combined to generate yet stronger constraints. If disagreement is found this is motivation for observational groups to collect more data or for theoretical groups to work out new mechanisms that can reconcile the discrepant observations.

Testing for statistical isotropy in the CMB is a well-defined operation, since it is easy to specify alternative models.  It requires checking whether a model of the correlations in the fluctuations  in terms of rotationally invariant quantities (such as the power spectrum) is better or worse than a model that contains quantities that do not transform as scalars under SO(3). A frequentist test statistic has been suggested in \cite{hajiansouradeep}, but a Bayesian treatment has not yet been attempted. The detection of a significant deviation from statistical isotropy would be a very important result, since isotropy is a fundamental prediction of inflation.

Testing for non-Gaussianity (NG) is similarly important, but much less well-defined. In the absence of physical models for NG a Bayesian treatment is not possible. The standard approach is to define some NG statistic, e.g. the skewness of the one-point function of the CMB fluctuations. Then this statistic is applied to the data and to a sample of Gaussian Monte Carlo samples (which are usually constrained to match the two-point statistics of the observed data). This Monte Carlo sample represents the null-hypothesis. A way is defined to assess discrepancy of the data and the Monte Carlo sample and if the discrepancy is statistically significant a detection is claimed.

Even though this is a straightforward frequentist procedure there is a great deal of arbitrariness in choosing the test statistic. Usually the choice is made based on a vague notion of genericity, in the sense that, for example, the skewness of the data is probably a more generic NG statistic than the temperature in pixel number 2,437,549, say. However, it is clearly not trivial to define a proper measure on the space of all possible statistics. The arbitrariness in the selection of the statistic (topological features of the map, n-point functions in pixel and in spherical harmonic space, wavelets, excursion sets, etc.) makes the test results difficult to interpret. For example, is clearly true that there will always be {\em some} statistic that gives an n-sigma result for any n!

To circumvent this arbitrariness, we could test for the NG predicted in certain variants of inflationary models (\cite{KomatsuSpergel,KSW,WMAPNG} and references therein). These physical models allow a Bayesian analysis in principle, and it would provide the best possible constraints on the presence of NG, but implementing Bayesian NG inference
these models is computationally tedious.

\section{Conclusion}
\label{conclusions}
In this talk I reviewed some of the more severe statistical and computational challenges of CMB analysis. CMB data are fundamentally important to cosmology, the problems I outlined are intellectually rich, and the knowledge that we need to make the most of the one CMB sky we can observe and analyze, present a powerful motivation to solve the CMB analysis problem.

\begin{acknowledgments}
I thank the organizers of
PHYSTAT2003 for making this conference so enjoyable
and stimulating. This work has been partially supported by the
National Computational Science Alliance under grant number
AST020003N and by the University of Illinois at Urbana-Champaign through an NCSA Faculty Fellowship.
\end{acknowledgments}


\end{document}